\documentclass[10pt,a4paper]{iopart}

\usepackage{iopams}
\usepackage{graphicx}
\usepackage{cite}
\usepackage{hyperref}
\usepackage[caption=false,labelformat=empty,farskip=0pt,topadjust=0pt,nearskip=0pt,captionskip=0pt]{subfig}
\usepackage{mathrsfs}

\hypersetup{colorlinks=true,citecolor=blue}

\newcommand{\text}[1]{\mathrm{#1}}
\newcommand{\neff}{n_\text{eff}}
\newcommand{\Lrad}{L_\text{rad}}
\newcommand{\Energy}{E}
\newcommand{\abs}[1]{\left| #1 \right|}
\newcommand{\micron}{{\mu\mathrm{m}}}
\newcommand{\Wcm}{\mathrm{Wcm}}
\newcommand{\MeV}{\mathrm{MeV}}

\begin{document}

\title{Nonlinear Breit-Wheeler pair creation with bremsstrahlung $\gamma$ rays}

\author{T. G. Blackburn and M. Marklund}
\address{Department of Physics, Chalmers University of Technology, SE-41296 Gothenburg, Sweden}
\ead{tom.blackburn@chalmers.se}

\date{\today}

\begin{abstract}
Electron-positron pairs are produced through the Breit-Wheeler process when energetic photons traverse electromagnetic fields of sufficient strength.
Here we consider a possible experimental geometry for observation of pair creation in the highly nonlinear regime, in which bremsstrahlung of an ultrarelativistic electron beam in a high-$Z$ target is used to produce $\gamma$ rays that collide with a counterpropagating laser pulse.
We show how the target thickness may be chosen to optimize the yield of Breit-Wheeler positrons, and verify our analytical predictions with simulations of the cascade in the material and in the laser pulse.
The electron beam energy and laser intensity required are well within the capability of today's high-intensity laser facilities.
\end{abstract}

\submitto{\PPCF}



\section{Introduction}

Breit-Wheeler pair creation is an elementary process of quantum
electrodynamics (QED) in which matter and antimatter are produced
purely from light~\cite{BreitWheeler}.
The two-photon, or linear, process has yet to be detected experimentally,
as it is difficult to achieve a collision between photon beams
where the flux is sufficiently high and the per-particle centre-of-mass
energy exceeds twice the electron mass.
Both these requirements have been met experimentally, and pair
creation observed, in the multiphoton regime: \cite{Burke} used
Compton scattering of a 46.6~GeV electron beam in a laser pulse
with strength parameter $a_0 = 0.36$ 
to produce $\gamma$ rays
that subsequently interacted with further laser photons to
produce electron-positron pairs~\cite{Bula}.

In this work we consider Breit-Wheeler pair creation in the
highly nonlinear regime, which is relevant for the study
of astrophysical plasmas in strong magnetic fields~\cite{Timokhin}
and is expected to occur prolifically in the next generation
of high-intensity laser experiments~\cite{BellKirk}.
Prospects are good for experimental exploration of this regime
with currently existing laser facilities, due to advances
in laser wakefield acceleration (LWFA)~\cite{Esarey} and
increases in available laser power. It is now
possible to accelerate electrons to multi-GeV energies in relatively compact
setups~\cite{Kim,Wang,Leemans} and to focus laser pulses to
intensities $> 10^{22}~\Wcm^{-2}$~\cite{Hercules,JKaren}.
Combining these lets us study the dynamics of energetic
particles in electromagnetic fields of unprecedented strength
using `all-optical' designs~\cite{BulanovDesign}.
Indeed, observation of radiation reaction (recoil due to
photon emission) in the collision of a LWFA electron beam
with an intense laser pulse has recently been reported~\cite{Cole,Poder}.

	\begin{figure}
	\centering
	\includegraphics[width=0.8\linewidth]{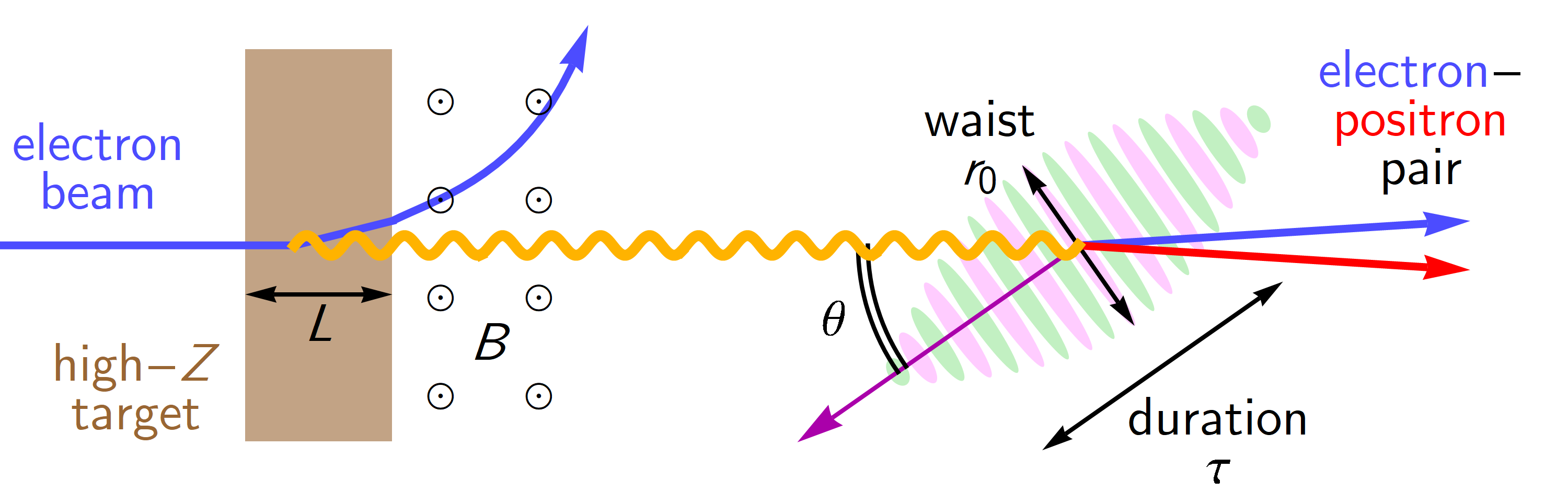}
	\caption[Diagram]
		{
		An ultrarelativistic electron beam, produced by laser wakefield
		acceleration (not shown), strikes a high-$Z$ target. The bremsstrahlung
		$\gamma$ rays so produced are separated from the charged components
		of the cascade by magnetic deflection, and collide downstream with
		an intense laser pulse. Here they produce electron-positron pairs
		via the nonlinear Breit-Wheeler process.}
	\label{fig:Diagram}
	\end{figure}

The configuration we study is the collision of GeV $\gamma$
rays with a laser pulse that has $a_0 > 10$.
A possible experimental realisation of this is illustrated
in \fref{fig:Diagram}, following \cite{EliTdr}.
The $\gamma$ rays are created by bremsstrahlung of a
LWFA electron beam in a high-$Z$ target; the ultrashort,
energetic $\gamma$ ray bunches this produces already find
applications in imaging and radioisotope generation
(see \cite{Ledingham,Albert} and references therein).
A gap is introduced between the solid target
and point of collision with the laser to permit magnetic
deflection of the source electrons and electron-positron
pairs produced inside the target, ensuring that we
have a pure light-by-light collision.

The importance of QED effects is measured by the parameter~\cite{Ritus}
	\begin{equation}
	\chi_\gamma = \frac{a_0 \omega_0 \omega (1 + \cos\theta)}{m^2},
	\end{equation}
where $\omega$ is the photon energy, $a_0$ and $\omega_0$
are the laser strength parameter and frequency, $\theta$
is the collision angle between the two (see \fref{fig:Diagram}),
and $m$ is the electron mass.
(Natural units $\hbar = c = 1$ are used throughout this paper.)
The onset of pair creation occurs for $\chi_\gamma \gtrsim 0.1$,
or when $(\omega/\mathrm{GeV})(a_0/20) \gtrsim 1$
at a wavelength of $0.8~\micron$.

Using bremsstrahlung to produce the seed photons, rather
than Compton scattering in a direct collision between
electron beam and laser pulse~\cite{Blackburn}, is
motivated by the breadth of the energy spectrum~\cite{Tsai}.
As it extends up to the initial energy of the electron,
using GeV electron beams will produce the GeV photons
that are necessary for $\chi_\gamma \gtrsim 0.1$.
Photons of this energy could also be
used to study the linear Breit-Wheeler process, either by colliding the
$\gamma$ rays with the high-temperature X-ray bath in
a laser-irradiated hohlraum~\cite{Pike}, or by colliding two
such beams directly~\cite{Ribeyre}. They could also
be used to seed QED avalanches at intensities
$> 10^{23}~\Wcm^{-2}$~\cite{Tang}.

By using a laser pulse with $a_0 > 10$ as the target,
we enter the strong-field regime where the pair creation
probability increases non-perturbatively with the laser
amplitude. This will permit the positron yield to be
substantially higher than reported by \cite{Burke}
despite the lower electron beam energies we consider.
To show this, we first calculate
an estimate for the pair creation probability in
\sref{sec:PP} and then show that bremsstrahlung is
a good source of sufficiently energetic photons in
\sref{sec:Photons}. Then we combine these two to
estimate the number of pairs per electron in
\sref{sec:Positrons}. We find that the thickness of
the high-$Z$ target may be chosen to maximize the
number of positrons that are produced in the laser pulse
and discuss the importance of reducing the divergence
of the $\gamma$ ray beam.




\section{Probability of Breit-Wheeler pair creation}
\label{sec:PP}

We begin by determining an analytical estimate for the probability that
an electron-positron pair is created when a photon with energy $\omega$
collides with a laser pulse that has $a_0 \gg 1$.
We employ the locally constant field approximation (LCFA)~\cite{Ritus}, using
probability rates that are calculated in an equivalent system of fields
in which the local value of $\chi$ is the same~\cite{Erber,BKS}. This requires
$a_0^3/\chi \gg 1$, as will always be the case here~\cite{Dinu}.
(See \cite{Meuren} and references therein for a discussion of how
the pair creation probability may be calculated exactly in the framework
of strong-field QED.)

While the $\chi$ parameter, which determines the importance of QED
effects, would be maximized for a head-on collision between photons
and laser pulse, we show in \fref{fig:Diagram} a crossing angle $\theta > 0$.
This is likely to be unavoidable in future laser experiments,
as it prevents damage to the focussing optics by transmitted light
and high-energy particles~\cite{EliTdr}.
It is necessary therefore to take the transverse structure of the focussed
laser pulse into account when calculating the positron yield, as the
distance over which the $\gamma$ rays are exposed to the strong fields
depends upon both the laser's temporal duration and focal spot size.

Recent studies of strong-field QED processes in focussing
laser fields include:
exact calculation from QED of the pair creation probability
for the head-on collision of a photon and a tightly-focussed
laser pulse~\cite{DiPiazza};
and determination of the intensity threshold for a pair cascade
to be launched by two counterpropagating, tightly focussed
laser pulses~\cite{Jirka}.
In both cases a description of the electromagnetic field that
goes beyond the paraxial approximation is used, e.g.~\cite{Salamin}.

While this captures the angular divergence of a tightly-focussed
laser pulse, the transverse intensity profile measured at the focal
plane is rarely so ideal~\cite{Samarin}.
To capture the essential physics in our analytical calculation,
we consider the laser pulse to be a `light bullet'
with Gaussian transverse intensity profile of constant size.
The duration of the pulse, defined as the full width at half maximum
(FWHM) of the temporal intensity profile, is given by $\tau$.
The radius of the beam is given by $r_0$, the distance over which
intensity falls to $1/e^2$ of its central value.
We expect that additional effects, such as the finite size of the
$\gamma$ ray beam and spatiotemporal offsets,
may be accounted for approximately by modifying the effective peak
amplitude $a_0$~\cite{BlackburnPPCF}.

The quantum nonlinearity parameter at time $t$ of a photon with energy
$\omega$ colliding with a linearly-polarized laser pulse with normalized amplitude $a_0$
and angular frequency $\omega_0$ at crossing angle $\theta$ is
	\begin{equation}
	\chi_\gamma =
		\frac{a_0 \omega_0 \omega (1 + \cos\theta)}{m^2}
		\abs{\sin\phi}
		\exp\!\left( -\frac{\ln{2}\,\phi^2}{2\pi^2 \neff^2} \right)
	\label{eq:PhotonChi}
	\end{equation}
where $\phi = (1 + \cos\theta)\omega_0 t$ and
	\begin{equation}
	\neff =
		\frac{\omega_0 \tau}{2\pi}
		\left[
			1 + \frac{\tau^2 \tan^2(\theta/2)}{r_0^2 \ln{4}}
		\right]^{-1/2}
	\label{eq:Neff}
	\end{equation}
is the number of wavelengths that characterizes the effective pulse duration.

Integrating the probability rate for pair creation from \cite{Erber}
over the trajectory specified by \eref{eq:PhotonChi}, using the same
saddle-points method as \cite{Blackburn}, we find that the pair creation
probability
	\begin{equation}
	P_\pm \simeq
		\alpha a_0 \neff
		\mathcal{R}\!\left[ \frac{a_0 \omega_0 \omega  (1 + \cos\theta)}{m^2} \right]
	\label{eq:PairProbability}
	\end{equation}
where $\alpha$ is the fine-structure constant, $\neff$ is as given in \eref{eq:Neff}
and
	\begin{equation}
	\mathcal{R}(x) = \frac{0.453 K_{1/3}^2(\frac{4}{3x})}{1 + 0.145 x^{1/4} \ln(1+2.26x) + 0.330 x}
	\end{equation}
as in \cite{Blackburn}. The argument of $\mathcal{R}$ in \eref{eq:PairProbability} is the peak $\chi_\gamma$ of the photon.

	\begin{figure}
	\centering	
	\subfloat[]{\label{fig:PP-a}\includegraphics[width=0.8\linewidth]{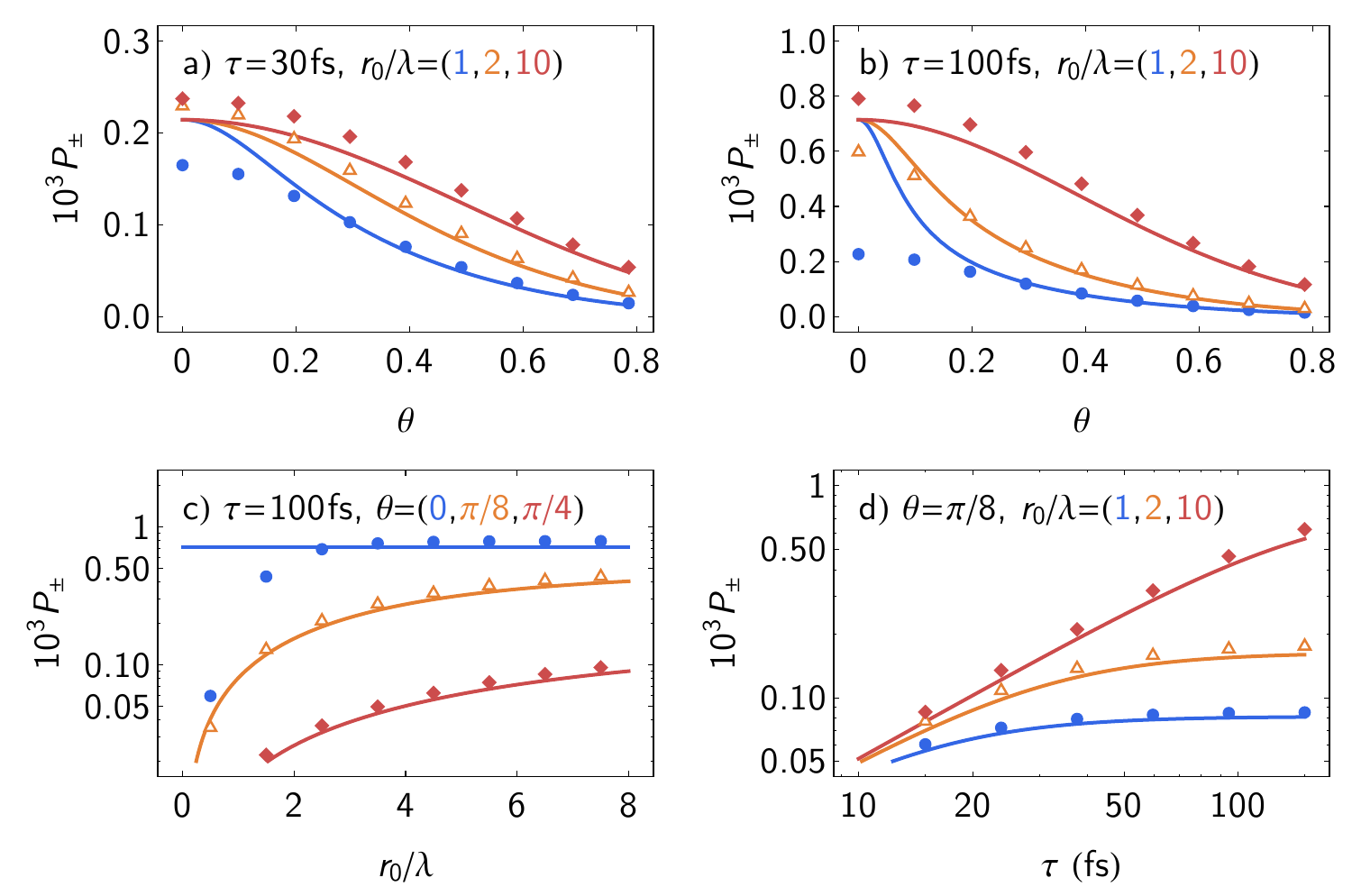}}%
	\subfloat[]{\label{fig:PP-b}}%
	\subfloat[]{\label{fig:PP-c}}%
	\subfloat[]{\label{fig:PP-d}}%
	\caption[Pair creation probability]%
		{The probability of pair creation $P_\pm$ in the collision of a
		$\gamma$~ray with energy $\omega = 1000 m$ and laser pulse
		with $a_0 = 30$ and wavelength $0.8~\micron$, as a function
		of the crossing angle $\theta$, the laser (FWHM) duration
		$\tau$ and focal spot size $r_0$: (lines) from
		\eref{eq:PairProbability} and (points) numerical
		integration.
		}
	\label{fig:PairProbability}
	\end{figure}

This analytical scaling may be verified against numerical integration of the
pair creation rate. In the latter we explicitly account for the effects of tight focussing
and model the spatial dependence of the pulse as
a Gaussian beam of spot size $r_0$ and Rayleigh range $z_R = \pi r_0^2 / \lambda$.
The fields are calculated to fourth-order in the diffraction angle $\epsilon = r_0/z_R$,
i.e. beyond the paraxial approximation~\cite{Salamin}. The temporal envelope
of the pulse remains a Gaussian with FWHM duration $\tau$.
For definiteness, we fix the $\gamma$-ray energy $\omega = 1000 m$ and
the laser $a_0 = 30$ at a wavelength $\lambda = 0.8\,\micron$, which corresponds
to a peak intensity of $2\times10^{21}\,\Wcm^{-2}$.
The pair creation probability predicted by \eref{eq:PairProbability} is compared
to the numerical results for varying collision angle $\theta$, pulse duration $\tau$ and
focal spot size $r_0$ in \fref{fig:PairProbability}.

\Fref{fig:PP-a} and \fref{fig:PP-b} show that the pair creation probability is
maximized for a head-on collision and decreases with increasing collision angle.
This is because both $\chi_\gamma$ and $\neff$ are reduced for $\theta > 0$
(in the latter case, because $r_0 < \tau$).
The two effects may be separated by comparing the results for different
spot sizes: at $r_0 = 10\lambda$ (in red), the pulse is effectively a plane wave
and the decrease in $P_\pm$ is entirely due to the geometric dependence
of $\chi_\gamma$.

We expect that agreement between the analytical and numerical results should be
better for larger spot sizes, because we assumed a plane wave in deriving
\eref{eq:PairProbability}. Our scaling does
capture with good accuracy the dependence of the pair creation probability
on collision angle for the $2\lambda$ and $10\lambda$ spots.
However, for the case that $r_0 = \lambda$, there is good agreement only if
$\theta \gtrsim 0.2$. Otherwise the analytical result
overestimates $P_\pm(\theta=0)$ by 30\% if the pulse duration is 30~fs or 210\%
if it is 100~fs.

This error arises when
the pulse duration $\tau$ becomes larger than the confocal parameter $2 z_R$,
as the probe photon can then `observe' the variation in intensity caused by the
contraction and expansion of the laser pulse as it passes through focus;
were $\tau \ll 2 z_R$ instead, this variation would be small compared to that
of the pulse temporal envelope.

We can therefore place a limit on the validity of \eref{eq:PairProbability}
in terms of the effective number of cycles $\neff$ (defined by \eref{eq:Neff}):
	\begin{equation}
	\neff < 2\pi (r_0/\lambda)^2
	\label{eq:NeffCondition}
	\end{equation}
Alternatively, this may be expressed in terms of a minimum angle:
	\begin{eqnarray}
	\theta &> \theta_\mathrm{min}, \quad
	\tan^2\left(\frac{\theta_\mathrm{min}}{2}\right) &= \ln4 \left[ \left(\frac{\lambda}{2\pi r_0}\right)^2 - \left(\frac{r_0}{\tau}\right)^2 \right].
	\label{eq:AngleCondition}
	\end{eqnarray}
Evaluating this for $\tau = 100$~fs, we find that
$\theta_\mathrm{min} = (0.36, 0.14, 0)$ 
for $r_0/\lambda = (1,2,10)$ respectively.
Inspection of \fref{fig:PP-b} shows that these bounds
are consistent with the minimum angles for which our analytical scaling
agrees with the numerical results.

We may further use \eref{eq:AngleCondition} to determine the smallest
spot size at given collision angle for which our analytical scaling is valid.
For the range of angles $\theta = (0,\pi/8,\pi/4)$, we find
$r_0/\lambda = (2.4,0.93,0.45)$. 
This is in good agreement with the results shown in \fref{fig:PP-c},
where we compare the pair creation probability as a function of
spot size at fixed pulse duration.

Finally, we show results with fixed $\theta = \pi/8$ and varying pulse
duration in \fref{fig:PP-d}. The minimum spot size at this collision
angle is $r_0/\lambda = 0.93$, so we find excellent agreement between
our analytical predictions and the
numerical results across the explored parameter range.

Having verified its accuracy, we are now in a position to apply our
analytical result to the case that the high-energy photons are
generated by bremsstrahlung, as shown in \fref{fig:Diagram}.


\section{Bremsstrahlung photon generation}
\label{sec:Photons}

The pair creation probability \eref{eq:PairProbability} is strongly
suppressed for $\chi_\gamma \ll 1$. Reaching $\chi_\gamma \sim 1$
with the intensities that may be reached
with today's high-intensity lasers ($\sim 10^{21}\,\Wcm^{-2}$),
requires photon energies in the GeV range~\cite{Blackburn}.
We now turn to how bremsstrahlung of an ultrarelativistic electron beam
in a high-$Z$ material may be used as the source of such photons.
In particular, we will use our analytical results to show how the
bremsstrahlung process may be optimized to produce the greatest number
of Breit-Wheeler positrons.

For the ultrarelativistic particles under consideration here, the two processes that
dominate the evolution of an electromagnetic cascade within the material are bremsstrahlung photon
emission and Bethe-Heitler pair creation. These occur when electrons (or positrons)
and photons respectively interact with the Coulomb fields of individual heavy atoms.
To a good approximation, the effect of the material properties, such as atomic
number $Z$ and mass density $\rho$, on the bremsstrahlung
spectrum may be parametrized by using only its radiation length $\Lrad$.

	\begin{figure}
	\centering
	\subfloat[]{\label{fig:PS-a}\includegraphics[width=0.8\linewidth]{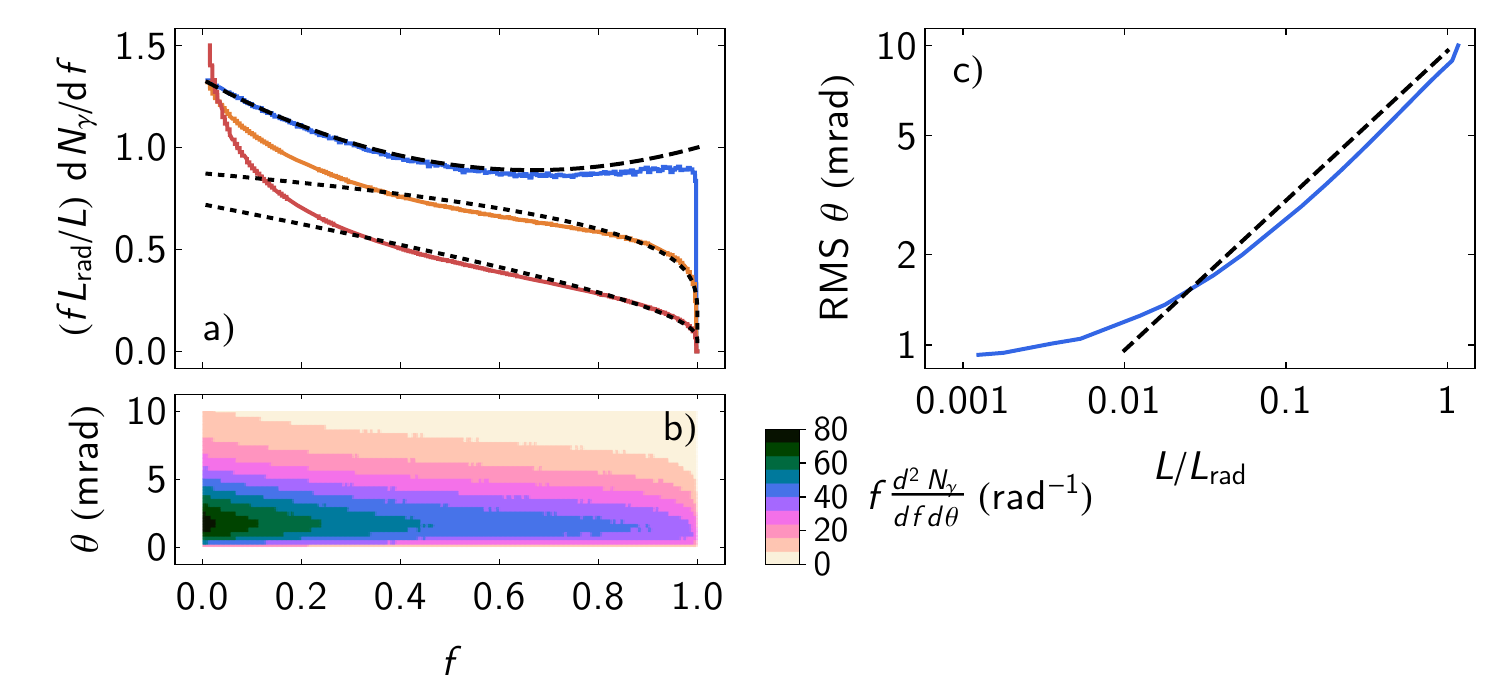}}%
	\subfloat[]{\label{fig:PS-b}}%
	\subfloat[]{\label{fig:PS-c}}%
	\subfloat[]{\label{fig:PS-d}}%
	\caption[Bremsstrahlung photon spectra]
		{Bremsstrahlung photon generation when a 2~GeV electron beam
		strikes a lead target of thickness $L$:
		a) energy spectra from (solid) simulations,
		(dashed) \eref{eq:ThinTargetSpectrum} and
		(dotted) \eref{eq:ThickTargetSpectrum}
		for $L = 0.2$~mm
		(blue), 2~mm (orange) and 5~mm (red);
		b) energy-divergence spectrum for $L = 2$~mm;
		c) the root-mean-square divergence of photons with $f>0.5$
		from (solid) \textsc{Geant4} simulations and
		(dashed) \eref{eq:PhotonDivergence}.}
	\label{fig:PhotonSpectra}
	\end{figure}

Under the approximations of complete screening and vanishing target thickness,
the number of photons produced with fractional energy $f = \omega/\Energy_0$ is
	\begin{equation}
	\frac{\rmd N_\gamma}{\rmd f} \simeq
		\frac{\ell}{f} \left( \frac{4}{3} - \frac{4f}{3} + f^2 \right)
	\label{eq:ThinTargetSpectrum}
	\end{equation}
where $\Energy_0 \gg m$ is the initial energy of the electron and $\ell = L/\Lrad$, the target
thickness $L$ scaled by the radiation length~\cite{BetheHeitler}.
\Eref{eq:ThinTargetSpectrum} neglects attenuation of the photon beam due to pair
creation within the solid target, thereby overestimating the high-energy tail of
the spectrum even for $\ell \simeq 0.01$.
This is particularly important here because the contribution to the Breit-Wheeler
positron yield will be dominated by the highest-energy photons.
A good approximation to attenuated bremsstrahlung spectrum for target thicknesses
$0.5 \lesssim \ell \lesssim 2$ is given by~\cite{Tsai}	
	\begin{equation}
	\frac{\rmd N_\gamma}{\rmd f} \simeq
		\frac{(1-f)^{4\ell/3} - e^{-7\ell/9}}{f \left[ \frac{7}{9} + \frac{4}{3}\ln(1-f) \right]}.
	\label{eq:ThickTargetSpectrum}
	\end{equation}

We compare \eref{eq:ThinTargetSpectrum} and \eref{eq:ThickTargetSpectrum} to the
results of \textsc{Geant4} simulations~\cite{Geant2003,Geant2006,Geant2016} for
electrons with $\Energy_0 = 2~$GeV striking lead targets of various thicknesses in
\fref{fig:PS-a}. (The radiation length of lead $\Lrad = 5.6$~mm.)
The spectra are broad, with substantial emission of photons with energy
greater than 1~GeV.
While the general shape of the spectrum at $L=0.2$~mm is captured well by
\eref{eq:ThinTargetSpectrum}, it is not very accurate near $f \simeq 1$.
For thicker targets, \eref{eq:ThickTargetSpectrum} gives better predictions
in the range $f > 0.5$, particularly for photons near the bremsstrahlung tip.
This will prove significant when we estimate the positron yield analytically,
as this is dominated by the highest-energy photons.

Due to the ultrarelativistic nature of the incident electrons, the emitted photons are
well-collimated around the forward direction: \fref{fig:PS-b} shows that for
$L = 2$~mm ($\ell = 0.36$) the typical divergence is 5~mrad and narrows slightly
with increasing photon energy.
Relativistic beaming means that we expect the divergence of the bremsstrahlung photons
to be inversely proportional to the Lorentz factor of the electron beam:
for $\ell \sim 1$, the root-mean-square (RMS) angle is approximately~\cite{SarriExpt2}
	\begin{equation}
	\theta_\mathrm{RMS}
		\simeq \frac{\sqrt{\ell}}{[E_0 / (19.2~\MeV)]}.
	\label{eq:PhotonDivergence}
	\end{equation}
A comparison with simulation
results shown in \fref{fig:PS-c} shows that this scaling works reasonably well.

We now discuss the implications of these results for the generation of Breit-Wheeler pairs.
The divergence of the $\gamma$~ray beam will play an important role because
it means that the beam will undergo transverse broadening as it propagates
over the distance between the high-$Z$ target and the focal plane of the secondary
laser pulse (see \fref{fig:Diagram}).
This reduces the number of $\gamma$~rays that actually hit the region
of highest intensity and so the positron yield.
(It would also alter the pair creation probability even for those photons that do hit the pulse, as $\chi_\gamma$ depends on $\theta$. However, for milliradian-level shifts, this is a relatively small effect compared to that of the reduced overlap.)

Assuming a divergence given by \eref{eq:PhotonDivergence}, the fraction of
photons $R$ that hit the focal spot (size $r_0$) after propagating a 
distance $D$ may be estimated as
	\begin{equation}
	R \simeq
		3\times10^{-5}
		\frac{(E_0/\mathrm{GeV})^2 (r_0/\micron)^2}{\ell (D/\mathrm{cm})^2}.
	\label{eq:OverlapRatio}
	\end{equation}
The importance of this reduction becomes clear when we consider that
$P_\pm \sim 10^{-4}$ (at $\tau = 30$~fs, see \fref{fig:PairProbability}).
Estimating the number of $\gamma$ rays to be equal to the number of
electrons in the bunch, $N_\gamma \sim 10^9$, we find that \eref{eq:OverlapRatio}
reduces the positron yield from $10^5$ to only one.
A possible way to overcome this would be to focus the
electron beam with a quadrupole magnet before it strikes the heavy target,
compensating for the increase in divergence during development of the
cascade~\cite{EliTdr}, and the intrinsic divergence of the electrons
(a few mrad in size for laser wakefield acceleration~\cite{Esarey}).

The more positive result is that the photons produced in
bremsstrahlung are sufficiently hard that they can used to probe nonlinear
Breit-Wheeler pair creation. \Eref{eq:ThickTargetSpectrum} predicts
that the number of photons per electron with $f > 0.5$ is as large as
$N_\gamma/N_e \simeq 0.2$ for $\ell \simeq 1$. Thus for electron beam
energies in excess of a GeV, as are available from laser wakefield
acceleration, we can expect a large number of photons with $\chi_\gamma$
sufficient for pair creation to take place.

\section{The positron yield and optimal target thickness}
\label{sec:Positrons}

	\begin{figure}
	\centering
	\includegraphics[width=0.5\linewidth]{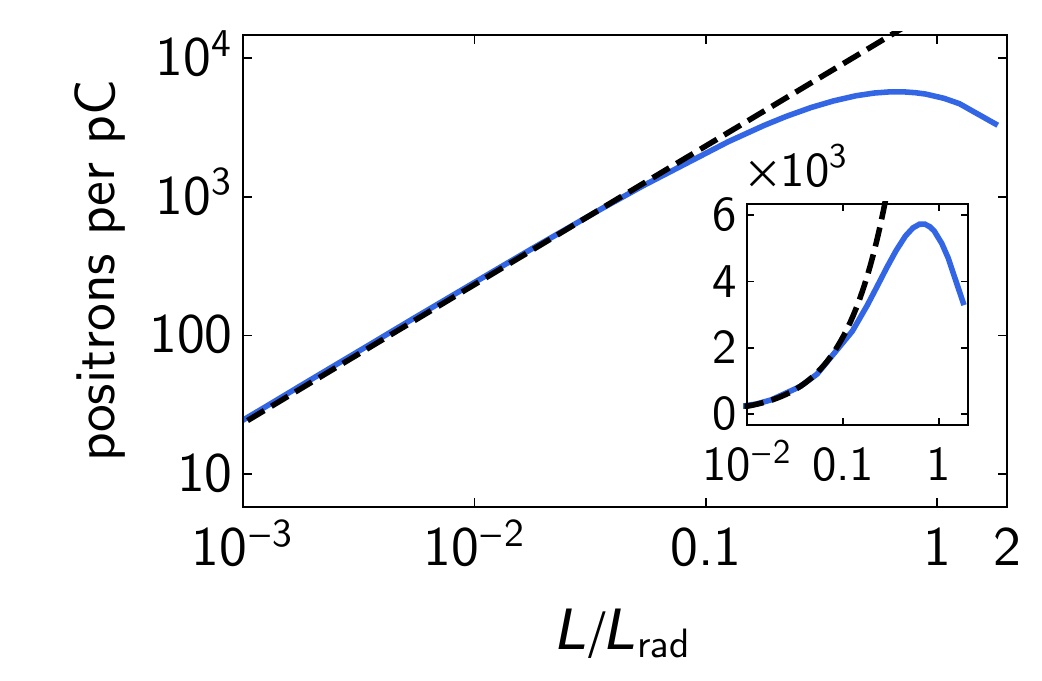}
	\caption[Positron yield]
		{The number of positrons per pC of charge in the
		electron beam,
		when the bremsstrahlung photons it produces in a lead target
		with thickness $L$ collide with a laser pulse
		that has $a_0 = 30$ and $\neff = 15$:
		(blue, solid) from simulations,
		(black, dashed) as predicted by \eref{eq:ThinTargetB}.
		}
	\label{fig:PositronYield}
	\end{figure}

The number of electron-positron pairs, per electron of the incident beam,
may be estimated by integrating the bremsstrahlung spectrum weighted by
the pair creation probability \eref{eq:PairProbability}:
	\begin{equation}
	N_\pm \simeq
		\alpha a_0 \neff \mathcal{B}(\ell, \chi_e).
	\label{eq:PositronYield}
	\end{equation}
Here we have defined an auxiliary function $\mathcal{B}$ to absorb the
positron yield's dependence on the properties of the solid target:
	\begin{equation}
	\mathcal{B}(\ell,\chi_e) =
		\int_0^1 \! \mathcal{R}(f \chi_e) \frac{\rmd N_\gamma}{\rmd f} \,\rmd f.
	\label{eq:B}
	\end{equation}
By using \eref{eq:ThinTargetSpectrum} or \eref{eq:ThickTargetSpectrum}
for $\rmd N_\gamma/\rmd f$, $\mathcal{B}$ becomes a function of only two
parameters: $\ell$, the scaled target thickness, and
$\chi_e = \Energy_0 a_0 \omega_0 (1+\cos\theta) / m^2$.
The former encapsulates the material properties through $\Lrad$.
The latter would be the quantum parameter of the electron, if it, rather
than its photons, collided with the laser pulse. It
depends upon the initial energy of the electron beam $\Energy_0$,
the normalized amplitude $a_0$ and angular frequency $\omega_0$ of the
intense laser pulse, and the crossing angle $\theta$ between the two.

Let us first consider the case where $\ell \ll 1$, so that we may use
\eref{eq:ThinTargetSpectrum} for the photon spectrum. It is evident
that the number of positrons increases linearly with target thickness,
as the number of bremsstrahlung photons does as well. In the limit
that $\chi_e \ll 1$, the integral in \eref{eq:B} may be performed analytically,
with the result $\mathcal{B} = \frac{3}{8} \ell \chi_e \mathcal{R}(\chi_e)$.
Otherwise, the integral must be performed numerically. A fit to these
results, accurate to 5\% over the range $0.01 < \chi_e < 10$ is
	\begin{equation}
	\mathcal{B}(\ell,\chi_e) \simeq
		\frac{0.375 \ell \chi_e \mathcal{R}(\chi_e)}{1 + 0.574 \chi_e^{2/3}}.
	\label{eq:ThinTargetB}
	\end{equation}
Considering that the prefactor $\alpha a_0 \neff \sim O(1)$ for
near-term experimental parameters, this may be used
for an order-of-magnitude estimate for the number of positrons
per electron. For $\ell \sim 0.1$ and $\chi_e \sim 1$, $\mathcal{B} \sim 10^{-3}$.
Thus if the bremsstrahlung photons from a few picocoulombs of
accelerated electrons reach the laser focal spot, we
can expect thousands of Breit-Wheeler positrons to be produced.

To verify this, we use \textsc{Geant4} to simulate the interaction of an
electron beam with a solid target, then take the resultant photon
spectrum as input to \textsc{Circe}~\cite{BlackburnPPCF,Blackburn},
a single-particle Monte Carlo code that
simulates strong-field QED cascades in intense laser pulses.
It does this by factorising the cascade into a product of first-order
processes (nonlinear Compton scattering and Breit-Wheeler pair
creation), which occur along the particle trajectory at locations
pseudorandomly determined according to the appropriate LCFA probability rate.
(See \cite{RidgersJCP,Gonoskov} for detailed discussion of this
``QED-PIC'' concept.)

We compare the positron yield predicted by \eref{eq:PositronYield}
and \eref{eq:ThinTargetB} and by simulations in \fref{fig:PositronYield}.
The electron beam energy is 2~GeV, and all the bremsstrahlung photons it
produces in a lead target collide head-on with
a laser pulse that has $a_0 = 30$, duration $\tau = 40$~fs and wavelength
$0.8~\micron$, i.e. $\neff = 15$.
We see that for $\ell < 0.1$, the positron yield increases linearly
with target thickness, in good agreement with \eref{eq:ThinTargetB}.
As $\ell$ continues to increase, the yield reaches a maximum of
$5\times10^3$ at $\ell = 0.7$ and then begins to decrease. This is
readily explained as the effect of pair creation within the solid
target, which causes attenuation of the high-energy part of the photon
spectrum. As it is these photons that are most likely to pair create,
increasing $\ell$ eventually causes the Breit-Wheeler positron yield
to decrease.

	\begin{figure}
	\centering
	\includegraphics[width=0.5\linewidth]{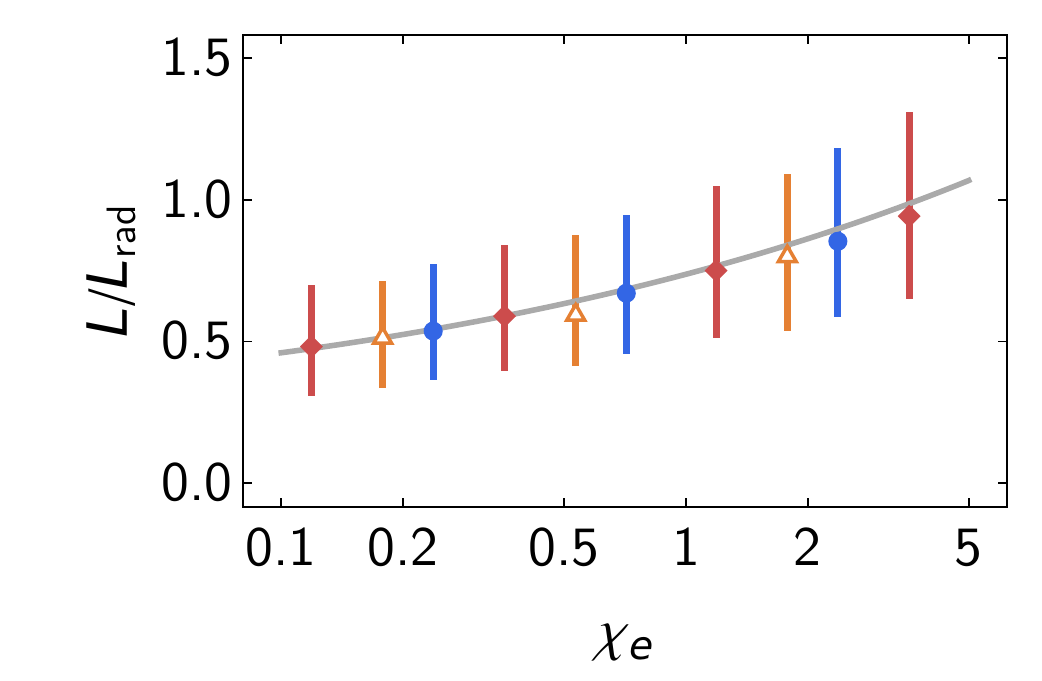}
	\caption[Optimal target thickness]
		{The target thickness $L$ per unit radiation length $\Lrad$
		which maximizes Breit-Wheeler pair creation when the bremsstrahlung
		photons collide with an intense laser pulse:
		(grey) as predicted by \eref{eq:OptimalEll}
		and (points) from simulations in which (blue) 2~GeV electrons
		hit a lead target, (orange) 1.5~GeV electrons hit copper
		and (red) 1~GeV electrons hit tantalum.
		Vertical bars indicate the range of $L/\Lrad$ over which the
		Breit-Wheeler positron yield is at least 95\% of maximum.
		}
	\label{fig:OptimalThickness}
	\end{figure}

If we use \eref{eq:ThickTargetSpectrum} rather than
\eref{eq:ThinTargetSpectrum} to model the photon spectrum, then we may
predict the $\ell$ which maximizes the yield of Breit-Wheeler
positrons. This is given by the root of the following integral equation:
	\begin{equation}
	\int_0^1 \!
	\mathcal{R}(f \chi_e)
	\frac{\partial^2 N_\gamma}{\partial\ell\partial f}
	\, \rmd f
		= 0,
	\label{eq:OptimalEll}
	\end{equation}
where the double differential photon spectrum is obtained from
\eref{eq:ThickTargetSpectrum}.
For convenience we solve this numerically for a range of $\chi_e$
and fit a two-component power law to the results.
We find that the optimal thickness
$\ell_\mathrm{opt} \simeq 0.693\chi_e^{1/4} + 0.0447\chi_e^{-1/5}$
for $0.01 < \chi_e < 10$, over which range the fit is accurate to
0.5\%.
As an example, if $\chi_e = 0.71$ as in \fref{fig:PositronYield},
\eref{eq:OptimalEll} predicts that the positron yield is maximized
at $\ell = 0.68$. This is in good agreement with the simulation
results, where we find $\ell_\mathrm{opt} = 0.67$.

Further verification of \eref{eq:OptimalEll} is shown in
\fref{fig:OptimalThickness}. Each point represents the optimal
$\ell$ found from a set of simulations in which the target thickness
is varied, while the electron beam energy $E_0$, target material,
and laser amplitude $a_0$ remain fixed. The materials under
consideration are lead, copper and tantalum, which have radiation
lengths of 5.61~mm, 14.4~mm and 4.09~mm respectively. The laser
$a_0$ is one of 10, 30 and 100 and $\neff = 15$ for all scans.
Our analytical prediction agrees well with the
simulation results across a broad range of electron beam and
target parameters.

We find that the optimal target thickness increases only slowly with increasing
$\chi_e$. Furthermore, the width of this maximum, indicated
by vertical lines in \fref{fig:OptimalThickness}, is large.
Therefore across the whole range $0.1 < \chi_e < 10$,
a positron yield close to maximum can be obtained simply
by setting $\ell \simeq 0.7$. This is well within expectations,
as the radiation length is approximately the distance
over which one photon or electron-positron pair
is added to the QED cascade in the material. Keeping $\ell \lesssim 1$
ensures that there are sufficient high-energy photons emitted
while minimising Bethe-Heitler pair creation.

This result further indicates that no special treatment is
required for electron beams with broad energy spectra,
i.e. a large spread in $\chi_e$. If $\ell \simeq 0.7$, the target
thickness is close to optimized for all components of the beam but
the low-energy tail, which contributes negligibly to pair creation.
Provided that there are picocoulombs of electrons with
$E_0 > 2$~GeV, then as shown in \fref{fig:PositronYield},
we expect thousands of positrons to be produced in a laser
pulse with $a_0 = 30$, i.e. a peak intensity of
$2\times10^{21}~\Wcm^{-2}$.

\section{Summary}

In this paper we have discussed the prospects for experimental
observation of nonlinear Breit-Wheeler pair creation, using
the collision between an intense laser pulse and
the $\gamma$ rays produced by bremsstrahlung of a LWFA
electron beam in a high-$Z$ target. 
We have shown that the thickness of the high-$Z$ target $L$
may be optimized to maximize the number of Breit-Wheeler
positrons: across a broad range of experimentally-accessible
parameters, this is $L/\Lrad = 0.7$, where $\Lrad$ is the radiation length.

However, we found that even though the divergence of the
$\gamma$ ray beam is small, it is sufficient to cause
most of the photons to miss the laser focal spot. This
due to transverse broadening of the $\gamma$ ray beam
as it traverses
the spatial separation between the solid target
and the focal plane of the laser pulse. (This separation is
required for magnetic deflection of the source electrons
and background electron-positron pairs.) As suggested in
\cite{EliTdr}, this makes it necessary to focus the
electron beam before it hits the high-$Z$ target.
Provided that this is done, the bremsstrahlung spectrum
of a multi-GeV electron beam with picocoulombs of charge
is sufficiently hard for thousands of positrons to be
produced in the intense laser pulse.


\ack
The authors
thank S. Yoffe and A. Noble for useful discussions and
acknowledge support from the Knut and Alice
Wallenberg Foundation and the Swedish Research Council
(grants 2013-4248 and 2016-03329).
Simulations were performed on resources provided by the Swedish National
Infrastructure for Computing (SNIC) at the High Performance Computing
Centre North (HPC2N).

\section*{References}
\bibliographystyle{iopart-num}
\bibliography{references}

\end{document}